\RequirePackage{rotating}
\documentclass[acmlarge,screen]{acmart}

\def\BibTeX{{\rm B\kern-.05em{\sc i\kern-.025em b}\kern-.08emT\kern-.1667em\lower.7ex\hbox{E}\kern-.125emX}}

\usepackage[utf8]{inputenc}
\usepackage[T1]{fontenc}
\usepackage{booktabs}
\usepackage{color, colortbl}
\usepackage{soul}
\soulregister\cite7
\soulregister\ref7
\soulregister\pageref7
\soulregister\url7
\soulregister\textit7
\usepackage{amsmath}
\usepackage{comment}
\usepackage{bm}
\usepackage{amssymb}
\usepackage{balance}
\usepackage{url}
\usepackage{xcolor}
\usepackage{rotating}
\usepackage{fontawesome}
\usepackage{pdfpages}
\usepackage{makecell}

\definecolor{lightyellow}{cmyk}{0,0,0.50,0}
\sethlcolor{lightyellow}
\definecolor{yellow}{cmyk}{0,0,0.50,0}

\definecolor{year}{HTML}{BEBADA}
\definecolor{concept}{HTML}{FB8072}
\definecolor{focus}{HTML}{80B1D3}
\definecolor{method}{HTML}{FDB462}
\definecolor{target}{HTML}{B3DE69}
\definecolor{tablehead}{HTML}{F2F2F2}

\usepackage{tikz}
\tikzset{every picture/.style={/utils/exec={\sffamily}}}
\usetikzlibrary{automata}
\usetikzlibrary{shapes.geometric,positioning}
\usetikzlibrary{matrix}
\usetikzlibrary{arrows.meta}
\usetikzlibrary{shapes}
\definecolor{light-gray}{gray}{0.90}
\pgfdeclarelayer{back}
\pgfsetlayers{back,main}
\def\drawfill#1;{
  \fill[light-gray] #1;
\begin{pgfonlayer}{back}
  \draw[line width=1pt] #1;
\end{pgfonlayer}}

\newcommand{\infobox}[4]{
    \begin{figure}[H]
        \centering
        \begin{tikzpicture}
            \node[anchor=text,text width=.9\columnwidth, draw, line width=1pt, fill=#3, inner sep=4mm] (big) {\\\small#4};
            \node[draw, line width=.5pt, fill=#2, anchor=west, xshift=5mm] (small) at (big.north west) {#1};
        \end{tikzpicture}
    \end{figure}
}

\usepackage{subcaption}
\usepackage{caption}

\acmJournal{CACM}

\begin{document}
\title{A Decade of Social Bot Detection}
\titlenote{Forthcoming article accepted for publication in \textit{Communications of the ACM}.}

\author{Stefano Cresci}
\email{stefano.cresci@iit.cnr.it}
\orcid{1234-5678-9012}
\affiliation{  \institution{Institute of Informatics and Telematics, National Research Council (IIT-CNR)}
  \streetaddress{via G. Moruzzi, 1}
  \city{Pisa}
  \state{Italy}
  \postcode{56124}
}

\renewcommand{\shortauthors}{S. Cresci}

\maketitle

\makeatletter{}On the morning of November 9th 2016, the world woke up to the shocking outcome of the US Presidential elections: Donald Trump was the 45th President of the United States of America. An unexpected event that still has tremendous consequences all over the world. Today, we know that a minority of social bots -- automated social media accounts mimicking humans -- played a central role in spreading divisive messages and disinformation, possibly contributing to Trump's victory~\cite{NBERw24631,grinberg2019fake}.

In the aftermath of the 2016 US elections, the world started to realize the gravity of widespread deception in social media. Following Trump's exploit, we witnessed to the emergence of a strident dissonance between the multitude of efforts for detecting and removing bots, and the increasing effects that these malicious actors seem to have on our societies~\cite{shao2018spread,stella2018bots}. This paradox opens a burning question: \textit{What strategies should we enforce in order to stop this social bot pandemic?} In these times -- during the run-up to the 2020 US elections -- the question appears as more crucial than ever. Particularly so, also in light of the recent reported tampering of the electoral debate by thousands of AI-powered accounts\footnote{\url{https://about.fb.com/news/2019/12/removing-coordinated-inauthentic-behavior-from-georgia-vietnam-and-the-us/}}.

What stroke social, political and economic analysts after 2016 -- deception and automation -- has been however a matter of study for computer scientists since at least 2010. In this work, we briefly survey the first decade of research in social bot detection. Via a longitudinal analysis, we discuss the main trends of research in the fight against bots, the major results that were achieved, and the factors that make this never-ending battle so challenging. Capitalizing on lessons learned from our extensive analysis, we suggest possible innovations that could give us the upper hand against deception and manipulation. Studying a decade of endeavours at social bot detection can also inform strategies for detecting and mitigating the effects of other -- more recent -- forms of online deception, such as strategic information operations and political trolls.

\includepdf{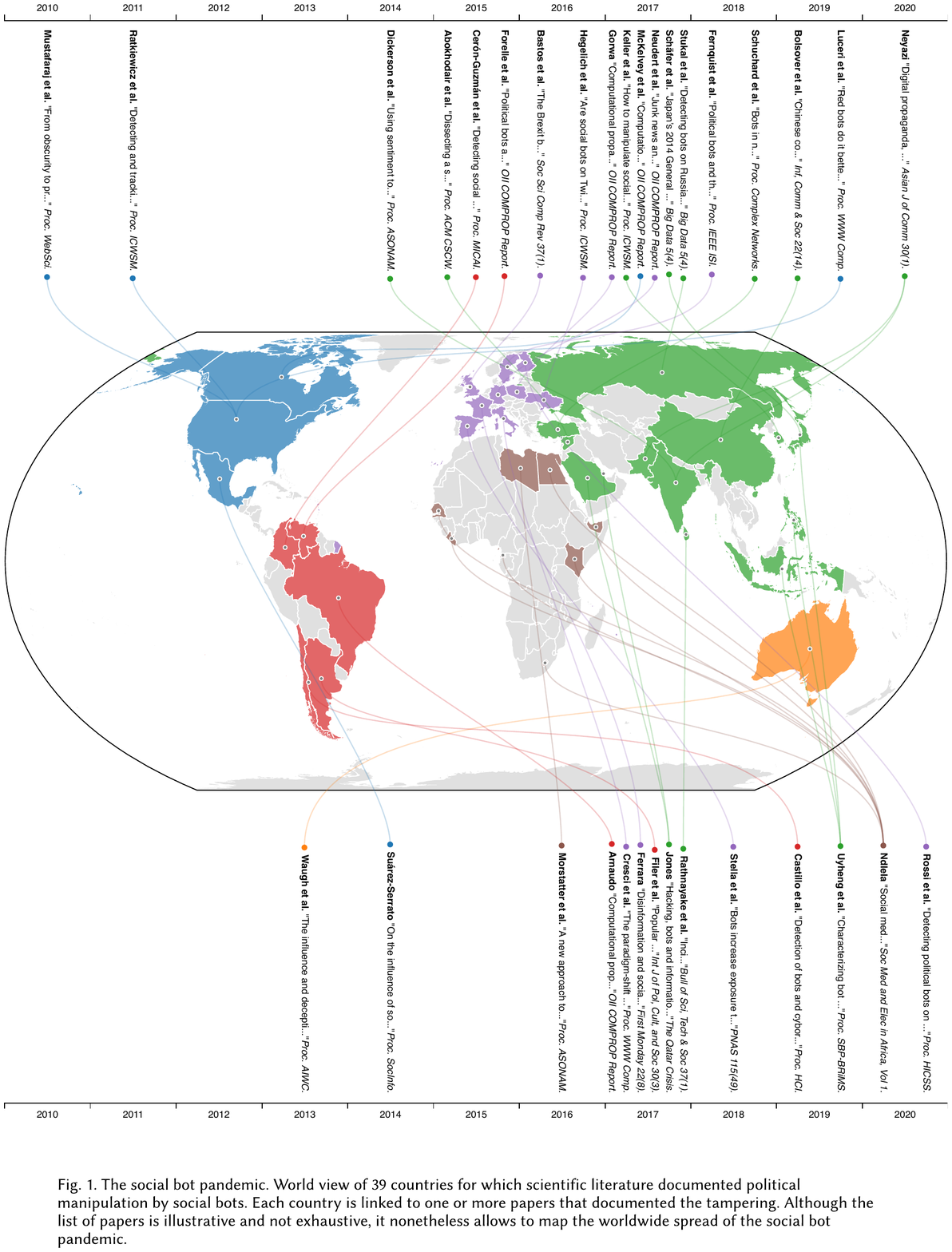}

\section*{The social bot pandemic}
Social bots coexist with humans since the early days of online social networks. Yet, we still lack a precise and well-agreed definition of what a social bot is. This is partly due to the multiple communities studying them and to the multifaceted and dynamic behavior of these entities, resulting in diverse definitions each focusing on different characteristics. Computer scientists and engineers tend to define bots from a technical perspective, focusing on features such as activity levels, complete or partial automation, use of algorithms and AI. The existence of accounts that are simultaneously driven by algorithms and by human intervention led to even more fine-grained definitions and cyborgs were introduced as either bot-assisted humans or human-assisted bots~\cite{chu2012detecting}. Instead, social scientists are typically more interested in the social or political implications of the use of bots, and define them accordingly. Social bots are actively used for both beneficial and nefarious purposes~\cite{ferrara2016rise}. Regarding the detection of benign or malicious social bots, the majority of existing works focused on detecting the latter. The reason is straightforward if we take into account the categorization proposed by Stieglitz \textit{et al.} in~\cite{stieglitz2017social}. Bots were categorized according to their \textit{intent} and to their capacity of \textit{imitating} humans, with the majority of existing specimen being either benign bots that do not aim to imitate humans (e.g., news and recruitment bots, bots used in emergencies) or malicious ones relentlessly trying to appear as human-operated. The detection of the former category of bots does not represent a challenge, and scholars devoted the majority of efforts to spot the latter, also because of their tampering with our online ecosystems. Indeed, the wide array of actions that social bots perform and the negligible cost for creating and managing them \textit{en masse}, open up the possibility to deploy armies of bots for information warfare, for artificially inflating the popularity of public characters and for manipulating opinions.

\setcounter{figure}{1}
\begin{figure*}[t]
    \centering
    \includegraphics[width=0.9\textwidth]{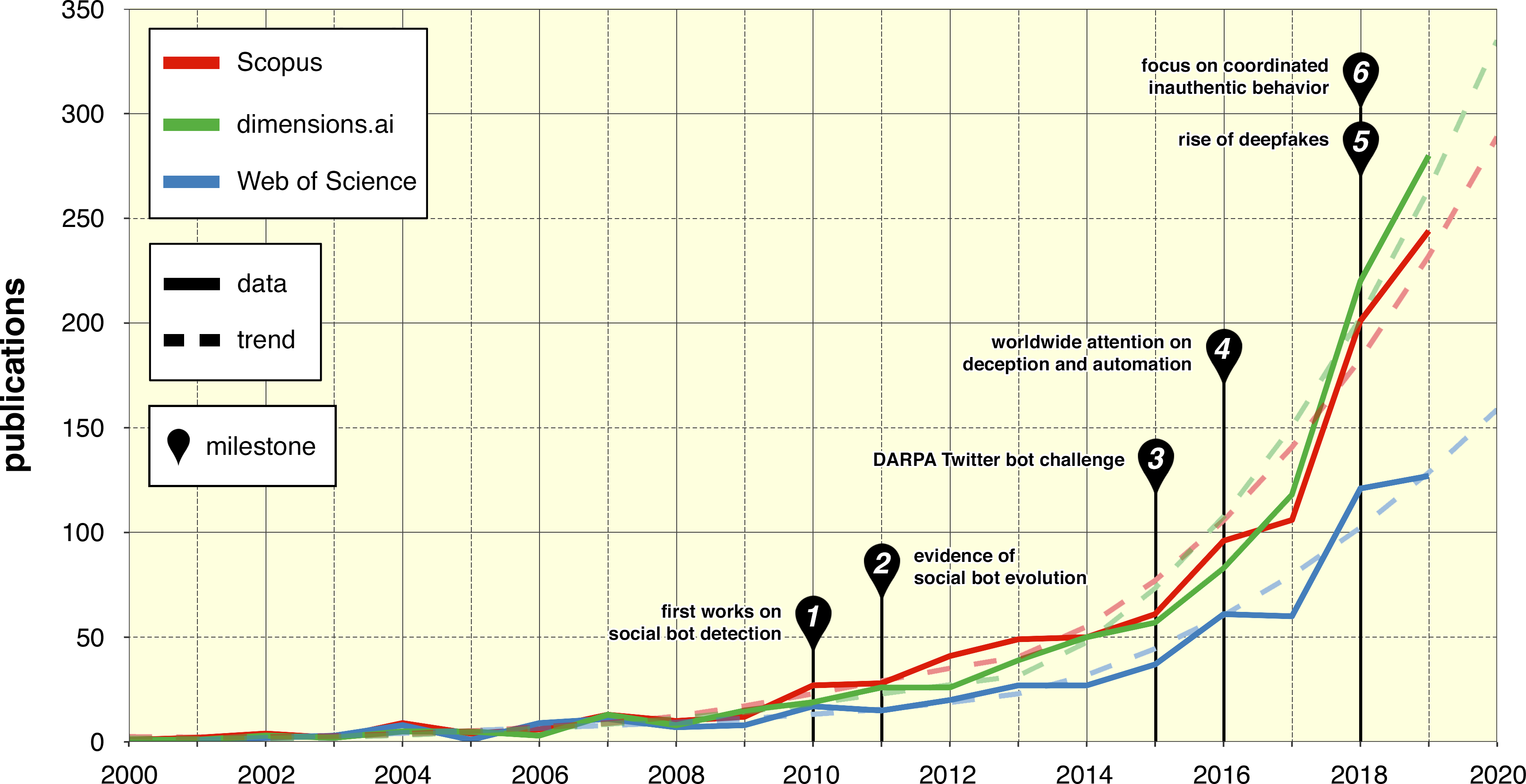}
    \caption{Publications per year on the characterization, detection and impact estimation of social bots. Since 2014, the number of publications on the topic skyrocketed. We forecast that from 2021 there will be more than 1 new paper published per day on social bots, which poses a heavy burden on those trying to keep pace with the evolution of this thriving field. Efforts aimed at reviewing and organizing this growing body of work are needed in order to capitalize on previous results.\label{fig:publications-trend}}
\end{figure*}

On the onset of the sudden surge of interest around automation and deception, several studies measured the extent of the social bot pandemic. Results are nothing less than worrying. The average presence of bots was estimated to be in the region of 15\% of all active Twitter accounts in 2017~\cite{varol2017online}, and 11\% of all Facebook accounts in 2019~\cite{zago2019screening} -- a considerable share indeed. Even more worryingly, when strong political or economic interests are at stake, the presence of bots dramatically increases. A 2019 study reported that 71\% of Twitter users mentioning trending US stocks, are likely to be bots~\cite{cresci2019cashtag}. Similar results were obtained about the presence of bots in online cryptocurrency discussions~\cite{nizzoli2020charting} and as part of the ``infodemics'' about the COVID-19 pandemic~\cite{gallotti2020assessing}. Other studies specifically focused on political activity, concluding that bots played a role in strategic information operations orchestrated ahead of numerous worldwide events, as shown in Figure~1. Despite taking part in political discussions about all countries highlighted in figure, bots did not always have a real impact. In fact, scholars still lack a widespread consensus on the impact of social bots, with some studies reporting on their pivotal role for increasing disinformation's spread, polarization, and hateful speech~\cite{stella2018bots,shao2018spread}, and competing results claiming that bots do not play a significant role in these processes~\cite{vosoughi2018spread}. The ubiquity of social bots is also partly fueled by the availability of open-source code, for which Bence Kollanyi reported an exponential growth that led in 2016 to more than 4,000 GitHub repositories containing code for deploying Twitter bots~\cite{kollanyi2016automation}. Other investigations demonstrated that this trend hasn't halted yet. In fact, by 2018, scholars found more than 40,000 public bot repositories~\cite{assenmacher2019openbots}. The looming picture is one where social bots are among the weapons of choice for deceiving and manipulating crowds. These results are backed by the same platforms where information operations took place -- namely, Facebook\footnote{\url{https://about.fb.com/news/2019/10/removing-more-coordinated-inauthentic-behavior-from-iran-and-russia/}}, Twitter\footnote{\url{https://about.twitter.com/en_us/values/elections-integrity.html}} and Reddit\footnote{\url{https://www.reddit.com/r/redditsecurity/comments/e74nml/suspected_campaign_from_russia_on_reddit/}} -- that banned tens of thousands accounts involved in coordinated activities since 2016.

Given the reported role of bots in several of the ailments that affect our online ecosystems, many techniques were proposed for their detection and removal -- adding to the large coverage that this topic also receives from news outlets -- contributing to the formation of a steeply rising publication trend.
Nowadays, new studies on the characterization, detection and impact estimation of bots are published at an impressive rate, as shown in Figure~\ref{fig:publications-trend}. Should this skyrocketing trend continue, by 2021 there will be more than one new paper published per day, which poses a heavy burden on those trying to keep pace with the evolution of this thriving field. Perhaps even more importantly, the rate at which new papers are published implies that a huge worldwide effort is taking place in order to stop the spread of the social bot pandemic. \textit{But where is all this effort leading?} To answer this question we first take a step back at the early days of social bot detection.

\makeatletter{}\section*{The dawn of social bot detection}

\begin{figure*}[t]
    \centering
    \includegraphics[width=0.75\textwidth]{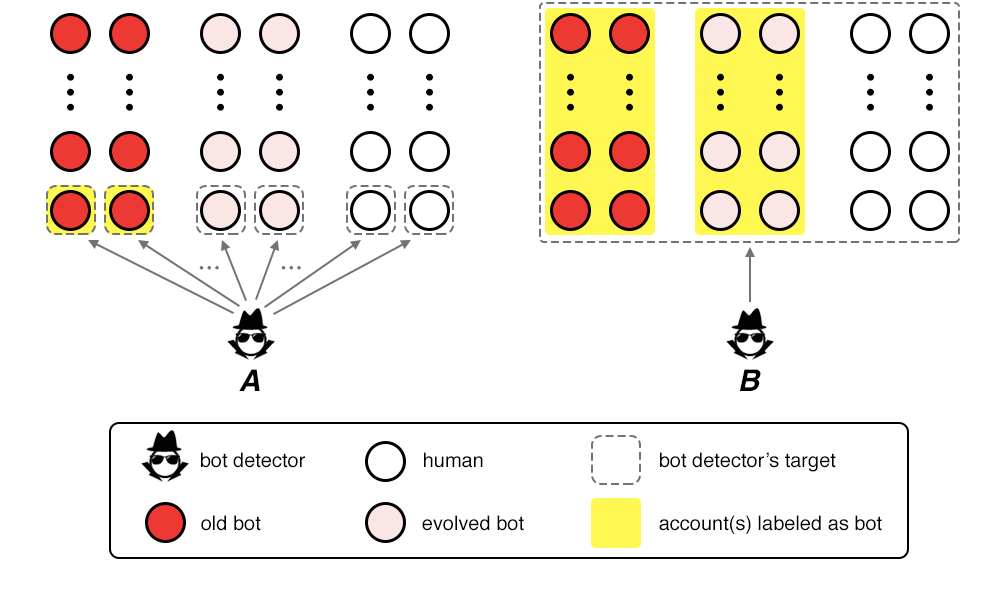}
    \caption[boh]{Differences between early and group approaches to social bot detection. In early approaches (panel \textbf{\textit{A}}), a supervised detector is separately applied to each account under investigation. If a bot does not appear as markedly different from a human-operated account, as in the case of recent evolved bots, it is likely to evade detection. In more recent approaches (\textbf{\textit{B}}), a detector analyzes a group of accounts, looking for traces of coordinated and synchronized behaviors. Large groups of coordinated accounts are more likely to be detected than sophisticated individual bots. Nonetheless, prediction errors can still occur for small groups of loosely coordinated bots that might provide insufficient information for detecting them, or for groups of highly coordinated humans that might appear as automated\protect\footnotemark. These issues currently represent unsolved challenges in the field.}\label{fig:bot-detection-approaches}
\end{figure*}

The first work that specifically addressed the detection of automated accounts in online social networks dates back to January 2010~\cite{yardi2010detecting}. In the early days, the vast majority of attempts at bot detection featured two distinctive characteristics: (i) they were based on supervised machine learning, and (ii) on the analysis of \textit{individual} accounts. In other words, given a group of accounts to analyze, detectors were separately applied to each account of the group, to which they assigned a binary label (either bot or legitimate). This approach to bot detection is schematized in panel \textbf{\textit{A}} of Figure~\ref{fig:bot-detection-approaches}. Here, the key assumption is that bots and humans are clearly separable and that each malicious account has individual features that make it distinguishable from legitimate ones. This approach to the task of social bot detection also revolves around the application of off-the-shelf, general-purpose classification algorithms on the accounts under investigation and on designing effective machine learning features for separating bots from legitimate accounts. 
For example, Cresci \textit{et al.} developed a set of supervised machine learning classifiers for detecting so-called fake followers, a type of automated accounts commonly used to artificially boost the popularity of the public characters that buy them~\cite{cresci2015fame}. Fake followers can be bought for as low as 12\$ per 1,000 followers in the surface Web. As a result, they are fairly common\footnote{\url{https://www.nytimes.com/interactive/2018/01/27/technology/social-media-bots.html}}. Cresci \textit{et al.} analyzed some 3 thousands fake followers obtained from different vendors and revealed that the simplistic nature of these accounts renders their detection rather easy, even when leveraging only 19 data- and computation-inexpensive features~\cite{cresci2015fame}. After all, fake followers need not perform complex tasks such as producing content or engaging in conversations. Other detection systems make use of large numbers of machine learning features to spot social bots. By leveraging more than 1,200 features of an account, Botometer evaluates possible bots based on their profile characteristics, social network structure, the content they produce, their sentiment expressions, and the timings of their actions~\cite{yang2019arming}. Instead of focusing on a specific type of bots, as Cresci \textit{et al.} did, Botometer represents a ``general purpose'' bot detector. The generality and ease of deployment of this detector are however counterbalanced by a reduced bot detection accuracy~\cite{cresci2017paradigm,grimme2018changing}. The two previous detectors simultaneously analyze multiple dimensions of suspicious accounts in order to spot possible bots. Instead, other systems solely focus on network characteristics, textual content of shared messages, or profile information. These systems are typically easier to game, since they only analyze a single facet of the complex behavior of bots.

\footnotetext{\url{https://ocean.sagepub.com/blog/devoted-users-eu-elections-and-gamification-on-twitter}}

Despite achieving promising initial results, these early approaches have a number of drawbacks. The first challenge in developing a supervised detector is related to the availability of a ground truth dataset to use in the training phase of the classifier. In most cases, a real ground truth is lacking and the labels are simply given by human operators that manually analyze the data. Critical issues arise as a consequence of the diverse definitions of social bots, resulting in different labeling schemes~\cite{grimme2017social}. Moreover, humans have been proven to suffer from several annotation biases and to largely fail at spotting recent sophisticated bots, with only 24\% bots correctly labeled as such by humans in a recent experiment~\cite{cresci2017paradigm}. Furthermore, these approaches typically output binary classifications. In many cases however, malicious accounts feature a mixture of automated and human-driven behaviors that cannot be accounted for with simple binary labels. To make matters worse, another major drawback of individual detectors is caused by the \textit{evolutionary} nature of social bots. 
\makeatletter{}\section*{The issue of bot evolution}
\begin{figure*}[t]
    \centering
    \includegraphics[width=\textwidth]{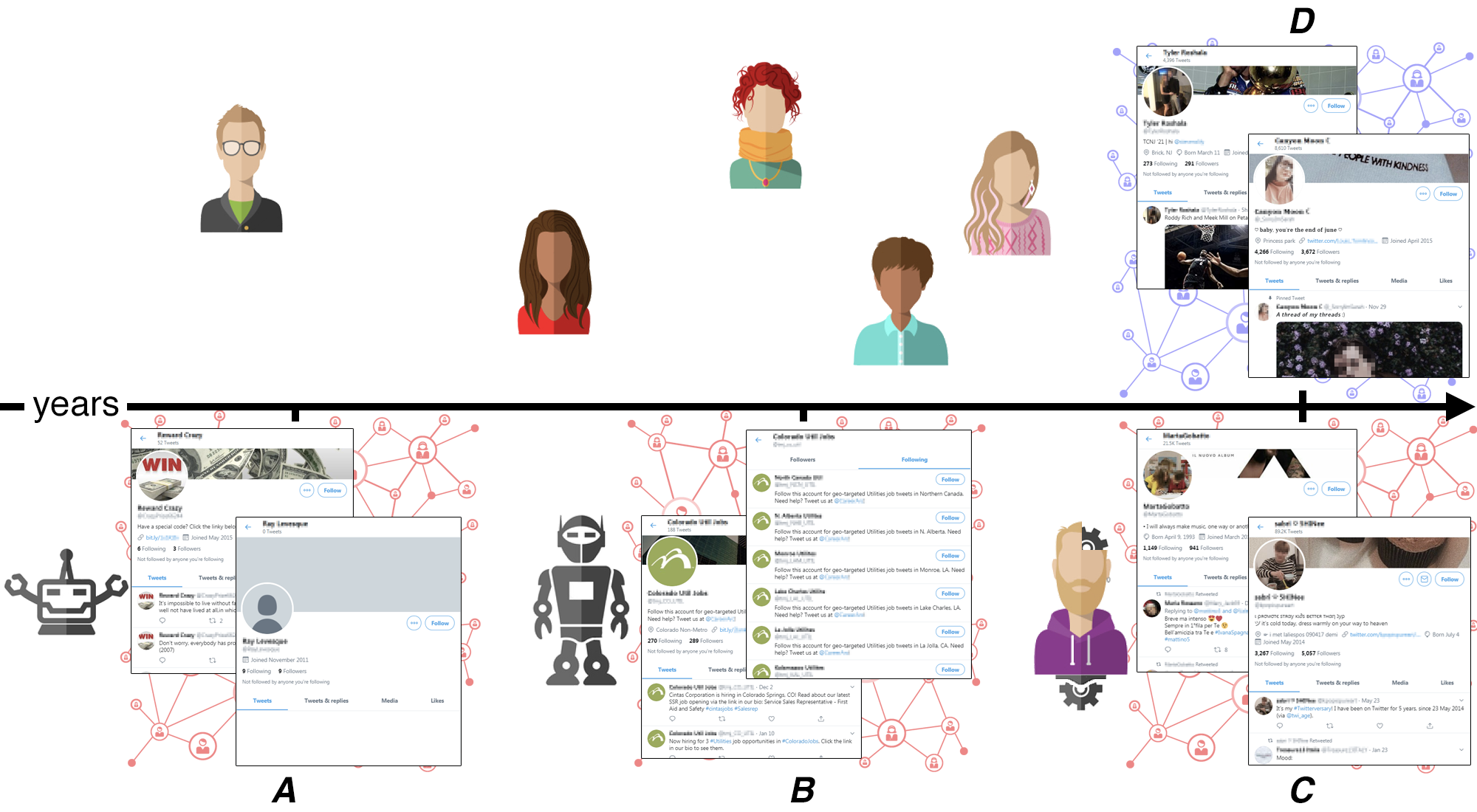}
    \caption{Example Twitter profiles showing the issue of bot evolution. Bots of the first wave (panel \textbf{\textit{A}}) were very simplistic, with few personal information and social connections. As such, they could be easily distinguished from human-operated legitimate accounts. The second wave consisted of more sophisticated accounts (panel \textbf{\textit{B}}), featuring detailed personal information. To increase their credibility, these bots often followed one another thus creating clearly identifiable botnets. Nowadays, social bots (panel \textbf{\textit{C}}) are so carefully engineered as to be more similar to human-operated accounts (panel \textbf{\textit{D}}) than to other bots. They have large numbers of real friends and followers, they use stolen names and profile pictures, and they intersperse few malicious messages with many neutral ones.\label{fig:bot-evolution}}
\end{figure*}
Initial success at social bot detection forced bot developers to put in place sophisticated countermeasures. Because of this, newer bots often feature advanced characteristics that make them way harder to detect with respect to older ones. This vicious circle leads to the development of always more sophisticated social bots and is commonly referred to as \textit{bot evolution}.

Noteworthy works published by Chao Yang \textit{et al.} between 2011 and 2013 provided the first evidence and the theoretical foundations to study social bot evolution~\cite{yang2013empirical}. The first wave of social bots that populated OSNs until around 2011 was made of rather simplistic bots. Accounts with very low reputation due to few social connections and posted messages, and featuring clear signs of automation as shown in panel \textbf{\textit{A}} of Figure~\ref{fig:bot-evolution}. Conversely, the social bots studied by Chao Yang \textit{et al.} appeared as more popular and credible, given the relatively large number of their social connections. In addition, they were no longer spamming the same messages over and over again. Leveraging these findings, authors developed a supervised classifier that was specifically designed for detecting evolving bots. Initially, the classifier proved capable of accurately detecting this second wave of bots. Time went by and new studies acknowledged the rise of a third wave of bots that spread through online social networks from 2016 onwards~\cite{ferrara2016rise,cresci2017paradigm}, as shown in panel \textbf{\textit{C}} of Figure~\ref{fig:bot-evolution}. Unfortunately, Yang's classifier for detecting evolving bots was no longer successful at spotting the third wave of malicious accounts~\cite{cresci2017social}.
The previous example serves as anecdotal evidence of bot evolution and of the detrimental effect it has on detectors. Additional quantitative evidence is reported in other studies that evaluated the \textit{survivability} of different bots -- that is, their capability of continually evading detection and avoiding being removed from social platforms -- and the ability of humans in spotting bots in-the-wild. Results showed that only 5\% of newer bots are removed from social platforms, whereas older ones are removed 60\% of the times~\cite{cresci2017paradigm}. Moreover, hundreds of tech-savvy social media users that participated in a crowdsourcing experiment were able to tell apart newer bots from legitimate users only 24\% of the times. The same users were instead able of spotting older bots 91\% of the times~\cite{cresci2017paradigm}.

The previous anecdotal and quantitative results tell us that current sophisticated bots are hardly distinguishable from legitimate accounts if analyzed one at a time, as supervised classifiers and crowdsourcing participants did. In fact, \textit{newer bots are more similar to legitimate human-operated accounts than to other older bots}. Among the reasons for the human-like appearance of many bots is an increased hybridization between automated and human-driven behaviors. These cyborgs exist and operate halfway between the traditional concepts of bots and humans, resulting in weakened distinctions and overlapping behaviors between the two. Moreover, they are now using the same technological weapons as their hunters, such as powerful AI techniques for generating credible texts (e.g., via the \texttt{GPT-2} and 3 deep learning models\footnote{\url{https://openai.com/blog/better-language-models/}}) and profile pictures (e.g., via \texttt{StyleGAN}s deep learning models\footnote{\url{https://www.wired.com/story/facebook-removes-accounts-ai-generated-photos/}}).
Indeed, the possibility for malicious accounts to leverage deepfake texts, profile pictures, and videos is worrying, and worthy of increased attention~\cite{dasanmartino2020survey}. Kate Starbird recently discussed a related issue in an inspiring piece on Nature~\cite{starbird2019disinformationspread}. Similarly to the hazy duality between ``bots'' and ``humans'', she posits that the boundaries between what is ``fake'' and what is ``real'', are blurring. To this end, human-like bots and cyborgs are just the tip of the iceberg, with other newer forms of deception -- such as political trolls and ``unwitting humans'' -- that are bound to make the online information landscape an even grimmer place. Figure~\ref{fig:bot-evolution} provides some examples of Twitter profiles that demonstrate how real-world bots evolved over the course of the years. As one form of ``social Web virus'', bots mutated thus becoming more resistant to our antibodies. The social bot pandemic gradually became much more difficult to stop. Within this global picture, dichotomous classifications -- such as human \textit{vs} bot, fake \textit{vs} real, coordinated \textit{vs} not coordinated -- might represent oversimplifications, unable to grasp the complexity of these phenomena and unlikely to yield accurate and actionable results.

Ultimately, the findings about the evolution of online automation and deception tell us that the na\"{i}ve assumption of early, supervised bot detection approaches -- according to which bots are clearly separable from legitimate accounts -- is no longer valid.

\makeatletter{}\section*{The rise of group approaches}
The difficulties at detecting sophisticated bots with early approaches rapidly gave rise to a new research trend. Since 2012-13, several different teams independently proposed new systems that, despite being based on different techniques and implementations, shared the same concepts and philosophy. As schematized in Figure~\ref{fig:bot-detection-approaches} (panel \textbf{\textit{B}}), the primary characteristic of these new systems, is that of targeting \textit{groups} of accounts as a whole, rather than individual accounts. The rationale for this design choice is that bots act in coordination with other bots, forming botnets to amplify their effects~\cite{zhang2016rise}. The existence of botnets does not necessarily imply that accounts are explicitly connected in the social network, but rather that they are maneuvered by a single entity and that they share common goals. As such, botnets leave behind more traces of their automation and coordination than those left behind by sophisticated single bots~\cite{cresci2017paradigm}. Devising techniques for spotting suspiciously coordinated and synchronized behaviors is thus likely to yield better results than analyzing individual accounts. In addition, by analyzing large groups of accounts, detectors also have access to more data for fueling powerful -- yet data-hungry -- AI algorithms. \textit{In 2018} -- roughly 5 years after the emergence of the group approach to bot detection -- \textit{also Facebook\footnote{\url{https://newsroom.fb.com/news/2018/12/inside-feed-coordinated-inauthentic-behavior/}} and Twitter\footnote{\url{https://help.twitter.com/en/rules-and-policies/platform-manipulation}} acknowledged the importance of focusing on ``coordinated inauthentic behaviors''}. The second common feature to the majority of group detectors is the proposal of important algorithmic contributions, thus shifting from general-purpose machine learning algorithms such as support vector machines and decision trees, to ad-hoc algorithms that are specifically designed for detecting bots, in an effort to boost detection performance. Finally, many group detectors are also based on unsupervised or semi-supervised approaches. Here the idea is to overcome the generalization deficiencies of supervised detectors that are severely limited by the availability of exhaustive and reliable training datasets~\cite{de2018lobo}.

\begin{figure*}
    \centering
    \begin{subfigure}{0.9\textwidth}
        \centering
        \includegraphics[width=1\linewidth]{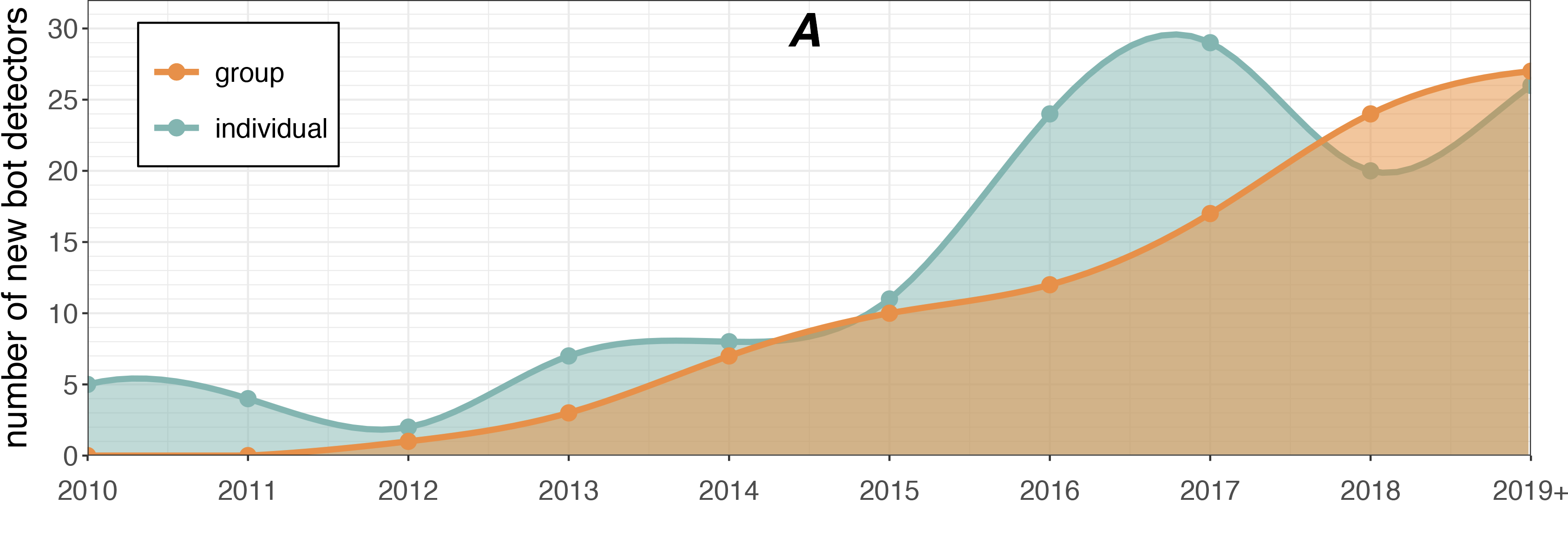}
    \end{subfigure}\\%
    \begin{subfigure}{0.9\textwidth}
        \centering
        \includegraphics[width=1\linewidth]{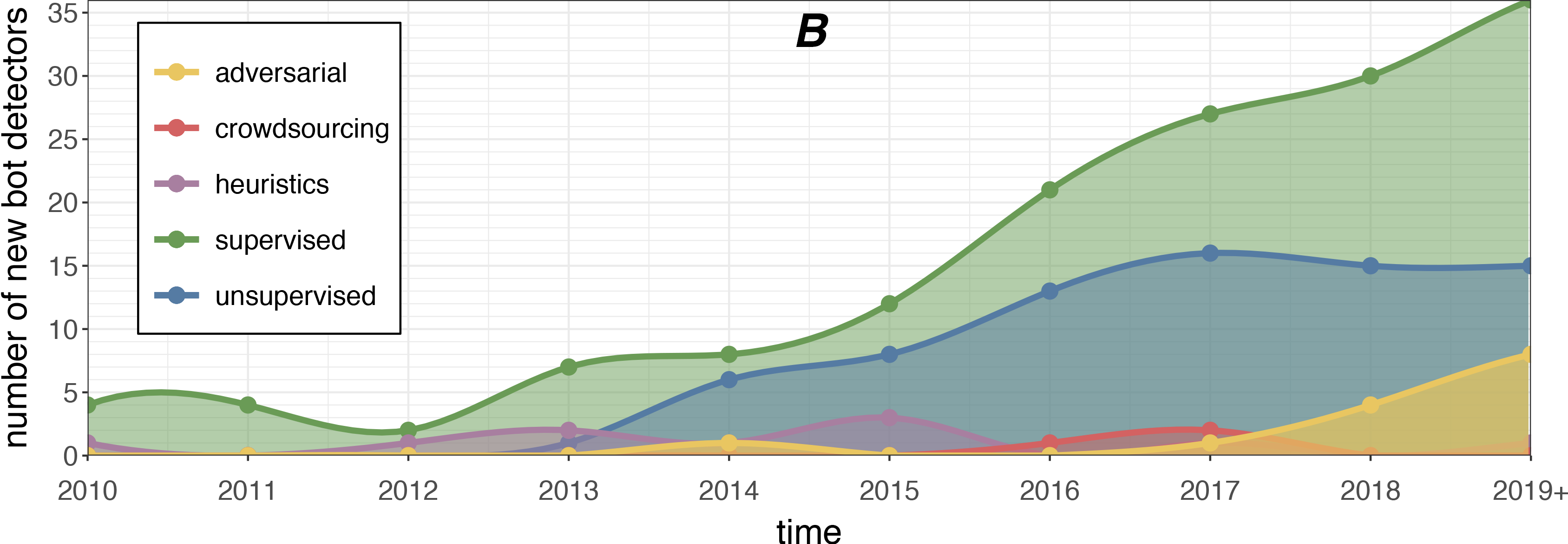}
    \end{subfigure}
\caption{Longitudinal categorization of 236 bot detectors published since 2010. Data points indicate the number of new detectors per type published in a given year. In panel \textbf{\textit{A}}, detectors are classified as either focusing on the analysis of individual accounts, or on the analysis of groups of accounts. In panel \textbf{\textit{B}}, the same detectors are classified based on their high-level approach to the task. Both panels clearly document the rise of a new approach to bot detection, characterized by group-analyses and many unsupervised detectors. Interestingly, the plateau reached by unsupervised approaches since 2017 occurred in conjunction with the recent rise of the adversarial ones.\label{fig:approaches-over-time}}
\end{figure*}

To quantitatively demonstrate the rise of group approaches to bot detection, in Figure~\ref{fig:approaches-over-time} we report the results of an extensive longitudinal classification. We surveyed more than 230 papers that proposed a bot detection technique and we manually classified each detector along two orthogonal dimensions. The first dimension (panel \textbf{\textit{A}}) highlights whether detectors target individual accounts or groups of accounts. Then in panel \textbf{\textit{B}}, we classify detectors according to their high-level approach to the task. In particular, we classified detectors as either based on: (i) heuristics -- i.e., based on simple rules; (ii) crowdsourcing -- that is, relying on the judgement of experts; (iii) supervised machine learning -- such as those based on classification and requiring a labeled training dataset; (iv) unsupervised machine learning -- such as those based on clustering that do not necessitate of labeled training data; or (v) adversarial approaches -- including adversarial machine learning. To better explain our methodology, in the following we briefly provide a couple of examples showing how well-known bot detectors were classified. The system proposed in~\cite{ruan2015profiling} is designed for detecting compromised accounts -- originally legitimate accounts that have been taken over by an
attacker. It initially builds a behavioral profile for each investigated account. Then, the system is able to detect compromised accounts via anomaly detection when a behavior diverges significantly with respect to its associated profile. This system is classified as based on the analysis of individual accounts (since the behavioral profile of an account depends solely on its own actions) and as an unsupervised detector (since it leverages an anomaly detection technique). Conversely, another system looks for suspiciously large similarities between the sequence of activities of vast groups of accounts~\cite{cresci2017social}. The activity of each account is encoded as a characters string and similarities between account activities are computed by applying the longest common subsequence metric to such strings. Suspiciously long subsequences between activity strings are identified via peak detection, and all those accounts that share the long activity subsequence are labeled as bots. Given these characteristics, this work contributes to group-based bot detectors (since it analyzes a group of accounts, looking for similar activity sequences) as well as to unsupervised machine learning approaches (since it leverages an unsupervised peak detection algorithm).
Generalizing the two previous examples, we note a few interesting patterns that derive from our classification. The vast majority of techniques that perform network analyses, for instance by considering the social or interactions graph of the accounts, are naturally classified as group-based. More often than not, they also propose unsupervised approaches. Contrarily, all techniques based on the analysis of the textual content of posted messages, such as those works that exclusively employ natural language processing techniques, are supervised detectors that analyze individual accounts. 

By leveraging classification results reported in Figure~\ref{fig:approaches-over-time}, we can also derive a number of additional insights. First of all, the rising publication trend of bot detectors follows the general trend of interest around social bots, previously shown in Figure~\ref{fig:publications-trend}. Indeed, since 2015 there has been a steadily increasing number of bot detectors published every year. From the trends shown in panel \textbf{\textit{A}} it is also strikingly evident that group-based approaches, revolving around the analysis of collective behaviors, are increasingly frequent. In fact, in 2018 the number of newly proposed group-based detectors surpassed for the first time that of detectors based on the analysis of individual accounts. From panel \textbf{\textit{B}} we note that bot detection approaches based on heuristics and crowdsourcing received very little attention. This is probably due to the many challenges involved in the development of these systems, which ultimately limit their applicability, scalability and detection performance. Instead, the number of new supervised detectors has been constantly increasing since 2012, despite their severe generalization issues~\cite{de2018lobo}. The adoption of unsupervised machine learning started in 2013 with the rise of group approaches, and now appears to be stationary. Interestingly, the plateau hit by unsupervised approaches co-occurred with the rise of adversarial ones, which might take their place in the coming years. Although the exact number of new bot detectors per type can slightly vary by analyzing a different set of papers, the big picture that emerges from Figure~\ref{fig:approaches-over-time} -- documenting the trends of \textit{individual}, \textit{group} and \textit{adversarial} approaches -- is clear, reliable and insightful. 
As a consequence of this paradigm-shift, group-based detectors are particularly effective at identifying evolving, coordinated, and synchronized accounts. For instance, several group detectors implement graph-based approaches and aim at spotting suspicious account connectivity patterns~\cite{jiang2016,nizzoli2020charting}. These techniques are suitable for studying both users interacting with content (e.g., retweets to someone else's tweets) or with other users (e.g., becoming followers of other users). Coordinated and synchronized behaviors appear as near-fully connected communities in graphs, dense blocks in adjacency matrices, or peculiar patterns in spectral subspaces~\cite{jiang2016inferring}. Other techniques adopted unsupervised approaches for spotting anomalous patterns in the temporal tweeting and retweeting behaviors of groups of accounts~\cite{chavoshi2016debot,mazza2019rtbust}. One way to spot accounts featuring suspiciously synchronized behaviors is by computing metrics of distance out of the accounts time series, and by subsequently clustering the accounts. The rationale behind this approach is based on evidence suggesting that human behaviors are intrinsically more heterogeneous than automated ones~\cite{cresci2019emergent}. Consequently, a large cluster of accounts with highly similar behaviors might indicate the presence of a botnet, even in the absence of explicit connections between the accounts. Distance between accounts time series was computed as a warp-correlation coefficient based on dynamic time warping~\cite{chavoshi2016debot}, or as the Euclidean distance between the feature vectors computed by an LSTM autoencoder~\cite{mazza2019rtbust}, a type of deep neural network that is particularly suitable for extracting latent features from sequential data.

As the switch from individual to group detectors demonstrates, the overall approach to the task of bot detection can have serious repercussions on detection performance. At the same time, some scientific communities tend to favor and stick to a specific approach. For instance, works published within the natural language processing community, quite naturally focus on textual content, thus resulting in a multitude of supervised classifiers that analyze accounts individually and that yield binary labels. In contrast, the complex networks community favors graph-based approaches. As a consequence, some combinations of approaches -- above all, text-based detectors that perform unsupervised, group analyses -- are almost unexplored and definitely underrepresented in the landscape of existing bot detectors. For the future, it would thus be advisable to multiply efforts along those directions that have been mostly overlooked until now. 
\makeatletter{}\section*{A glimpse into the future of deception detection}
So far, we highlighted that a shift took place from individual to group detectors in an effort to contrast social bot evolution. Now, we review the latest advances in the field for gaining possible insights into the future of deception detection. We ground this analysis on two observations.

Firstly, we observe that both the individual and the group-based approaches to social bot detection follow a reactive schema. In practice, when scholars and OSN administrators identify a new group of accounts that misbehave and that cannot be effectively detected with existing techniques, they react and begin the development of a new detection system. Hence, the driving factor for the development of new and better detectors have always been bot mischiefs. A major implication of this approach is that improvements in the detection of bad actors typically occur only some time after having collected evidence of new mischiefs. Bad actors such as bots, cyborgs and trolls thus benefit from a large time span -- the time needed to design, develop, and deploy a new effective detector -- during which they are essentially free to tamper with our online environments. In other words, \textit{scholars and OSN administrators are constantly one step behind of malicious account developers}. This lag between observations and countermeasures possibly explains the current situation with our online social ecosystems: Despite the increasing number of existing detection techniques, the influence of bots and other bad actors on our online discussions did not seem to decrease.

Our second observation is related to the use of machine learning for the task of social bot detection. The vast majority of machine learning algorithms are designed for operating within environments that are stationary and neutral, if not even benign. When the stationarity and neutrality assumptions are violated, algorithms yield unreliable predictions that result in dramatically decreased performances~\cite{goodfellow2018making}. Notably, \textit{the task of social bot detection is neither stationary nor neutral}. The stationarity assumption is violated by the mechanism of bot evolution that results in accounts exhibiting different behaviors and characteristics over time. Also the neutrality assumption is clearly violated, since bot developers are actively trying to fool detectors. As a consequence, the very same algorithms that we have been relying upon for a decade, and for which we reported excellent detection results in our studies, are actually seeing their chances to detect bots in-the-wild severely limited.  
Recent developments in machine learning may however come to our rescue and may possibly mitigate both issues emerging from the previous observations. Adversarial machine learning is a paradigm specifically devised for application in those scenarios presenting adversaries motivated in fooling learned models~\cite{goodfellow2018making}. Its high-level goal is to study vulnerabilities of existing systems and possible attacks to exploit them, before such vulnerabilities are effectively exploited by the adversaries. Early detection of vulnerabilities can in turn contribute to the development of more robust detection systems. One practical way to implement this vision is by generating and experimenting with adversarial examples -- that is, input instances specifically created to induce errors in machine learning systems.

\begin{figure*}[t]
    \centering
    \includegraphics[width=0.5\textwidth]{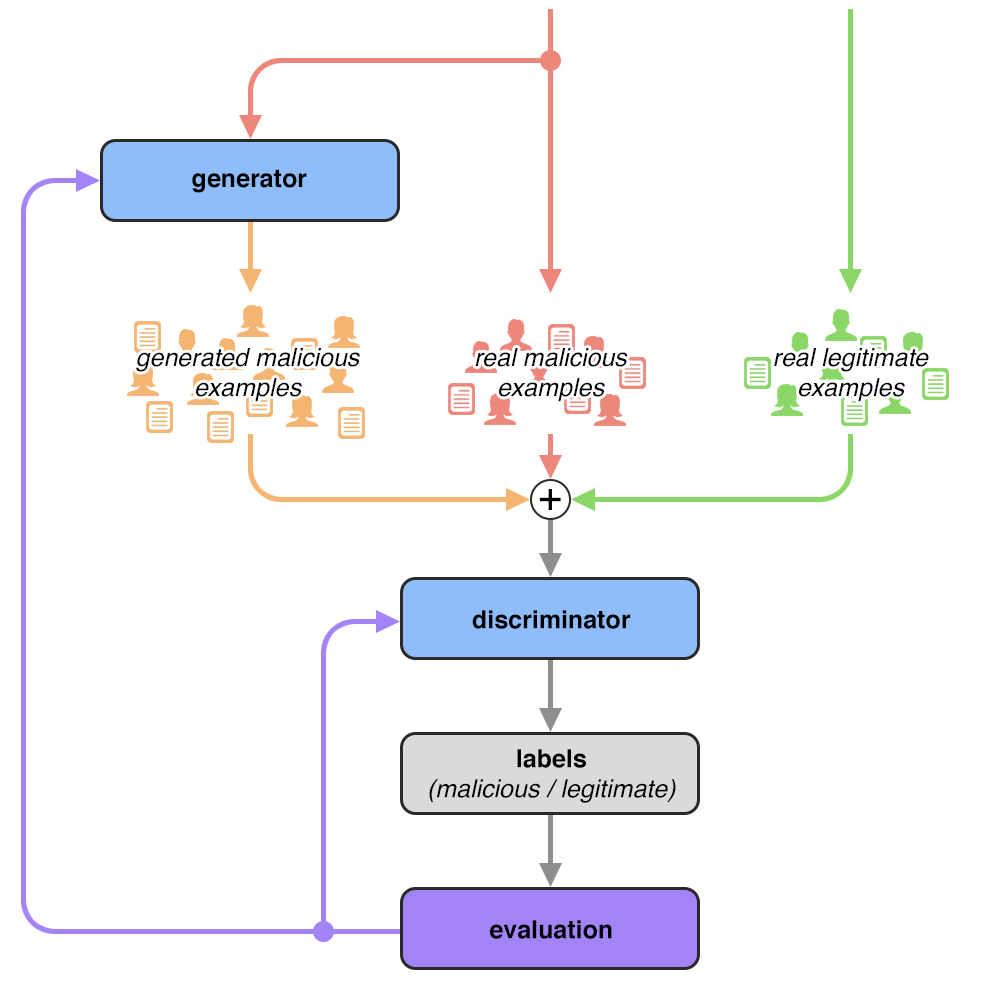}
    \caption{Adversarial deception detection based on generative adversarial networks (GANs). The generator network is employed for creating a large number of adversarial examples resembling the properties of real malicious examples. The discriminator network is trained to distinguish between malicious (either real or generated) and legitimate examples. By jointly training the two networks, the generator learns to produce more challenging malicious examples while the discriminator improves its overall classification performances since it trains on the challenging examples. This conceptual framework can be applied to many tasks, comprising the detection of disinformation, social bots and trolls.\label{fig:gan}}
\end{figure*}

\textit{All tasks related to the detection of online deception, manipulation and automation are intrinsically adversarial}. As such, they represent favorable application domains for adversarial machine learning. This intuition resulted in the first papers published in 2018-19 that initiated the development of an adversarial approach to bot detection, as shown in panel \textbf{\textit{B}} of Figure~\ref{fig:approaches-over-time}. In the so-called \textit{adversarial bot detection}, scholars experiment with meaningful adversarial examples with which they extensively test the capabilities of current bot detectors~\cite{cresci2019better}. Within this context, adversarial examples might be sophisticated types of existing bots and trolls that manage to evade detection by current techniques; or even bots that do not exist yet, but whose behaviors and characteristics are simulated, as done by Cresci \textit{et al.}~\cite{cresci2019better}; or bots developed ad-hoc for the sake of experimentation, as done by Grimme \textit{et al.}~\cite{grimme2018changing}. Finding good adversarial examples can help scholars understand the weaknesses of existing bot detection systems. As a result, bot hunters need not wait anymore for new bot mischiefs in order to adapt their techniques, but instead they can proactively (instead of reactively) test them, in an effort that could quickly make them more robust. In addition, this paradigm accounts for adversaries \textit{by design}, thus providing higher guarantees for deception detection, which violates the stationarity and neutrality assumptions.

The previous analysis highlights that initial efforts towards adversarial bot detection were driven by the creativity of some researchers and only covered few cases with limited applicability~\cite{grimme2018changing,cresci2019better}. In the near future they could instead be powered by the latest developments in AI. Generative adversarial networks (GANs) are a powerful machine learning framework where two competing deep learning networks are jointly trained in a game-theoretic setting~\cite{goodfellow2018making}. In particular, a GAN is composed of a \textit{generator} network that creates data instances and a \textit{discriminator} network that classifies data instances, combined as shown in Figure~\ref{fig:gan} where a GAN is instantiated for a generic task of deception detection. The goal of the generator is that of creating synthetic data instances that resemble the properties of real organic data, while the typical goal of the discriminator is to classify input data instances as either synthetic or organic. The discriminator is evaluated based on its binary classification performances, while the generator is evaluated in terms of its capacity to induce errors in the discriminator, hence the competition between the two networks. 
Originally, GANs were proposed as a form of generative model -- that is, the focus was posed on the generator network. A notable example of this kind is represented by the GAN trained in~\cite{wu2020using} for creating adversarial examples of social bots that improved the training of downstream detectors. However, with the end-goal of providing even larger improvements on deception detection, we could envision the adoption of GANs for training better discriminator networks. In particular, the generator of a GAN could be used as a generative model for creating many plausible adversarial examples, thus overcoming the previously mentioned limitations in this task and the scarcity of labeled datasets. Then, the whole GAN could be used to test the discriminator against the adversarial examples and to improve its detection performances. This paradigm has never been applied to the task of social bot detection, but it was tested with promising results for related tasks, such as that of fake news generation/detection~\cite{zellers2019defending}. The adversarial framework sketched in Figure~\ref{fig:gan} is general enough to be applied to a wide set of deception detection tasks, comprising the detection of social bots, cyborgs, trolls and mis/disinformation. Furthermore, in contrast with existing adversarial approaches for bot detection, it is grounded on an established and successful machine learning framework, rather than on ad-hoc solutions lacking broad applicability.

Despite the high hopes placed on adversarial approaches for detecting deception and automation, this research direction is still in its infancy and, probably due to its recency, is still lagging behind more traditional approaches. As such, efforts at adversarial detection can only be successful if the scientific community decides to rise to the many open challenges. Among them is the development of techniques for creating many different kinds of adversarial examples and to evaluate whether these examples are realistic and representative of future malicious accounts. In spite of these challenges, our analysis and the promising results obtained so far strongly motivate future endeavours in this direction, as also testified by the sparking adversarial trend in Figure~\ref{fig:approaches-over-time}.
 
\makeatletter{}\section*{Open challenges and the way ahead}
\begin{figure*}[t]
    \centering
    \resizebox{0.9\textwidth}{!}{\makeatletter{}\begin{tikzpicture}[
    mynode/.style={
	circle,
	draw=black,
	text opacity=1,
	inner sep=0pt,
	minimum size=12.5pt}]    \tikzstyle{axis} = [draw=black, -{Latex[length=3.75mm]}, line width=0.4mm]
    \tikzstyle{arrowline} = [draw=black, -{Latex[length=3.5mm]}, line width=0.2mm]
    \tikzstyle{simpleline} = [draw=black, line width=0.2mm]
    \tikzstyle{smallline} = [draw=black, -{Latex[length=2.5mm]}, line width=0.175mm]
            
        \draw [axis] (0,0) -- (0,7);
    \draw [axis] (0,0) -- (7,0);
        \node [align = right, font = \small] at (-0.4,1.5) {$b_0$};
    \node [align = right, font = \small] at (-0.65,5.5) {$b \ne b_0$};
    \node [align = center, font = \small] at (1.5,-0.4) {$t_0$};
    \node [align = center, font = \small] at (5.5,-0.4) {$t > t_0$};
        \node [align = center, rotate = 90] at (-1,3.5) {types of social bots ($b$)};
    \node [align = center] at (3.5,-1) {time ($t$)};
    
        \draw [simpleline] (-0.15,1.5) -- (1.5,1.5);
    \draw [simpleline, dashed] (1.5,1.5) -- (6,1.5);
    \draw [simpleline, dashed] (-0.15,5.5) -- (6,5.5);
    \node [align = center, font = \LARGE] at (6.75,1.45) {$\cdots$};
    \node [align = center, font = \LARGE] at (6.75,5.45) {$\cdots$};
    
        \draw [simpleline] (1.5,-0.15) -- (1.5,1.5);
    \draw [simpleline, dashed] (1.5,1.5) -- (1.5,6);
    \draw [simpleline, dashed] (5.5,-0.15) -- (5.5,6);
    \node [align = center, font = \LARGE] at (1.5,6.75) {$\vdots$};
    \node [align = center, font = \LARGE] at (5.5,6.75) {$\vdots$};
    
        \node[mynode,ultra thin,fill=target] at (1.5,1.5) {\Small\faCheck};
    \node[mynode,ultra thin,fill=method] at (1.5,5.5) {\small\faQuestion};
    \node[mynode,ultra thin,fill=method] at (5.5,1.5) {\small\faQuestion};
    \node[mynode,ultra thin,fill=concept] at (5.5,5.5) {\small\faQuestion};
    
        \def\legendmid{3.5}
    
        \node [align = center, font = \small] at (11,\legendmid+2.525) {generalize on:};
    \draw [smallline] (8.15,\legendmid+1.875) -- (8.65,\legendmid+1.875);
    \node [align = left, font = \scriptsize] at (9.4,\legendmid+1.875) {bot\\evolutions};
    \draw [smallline] (10.6,\legendmid+1.625) -- (10.6,\legendmid+2.125);
    \node [align = left, font = \scriptsize] at (11.4,\legendmid+1.875) {types\\of bots};
    \draw [smallline] (12.35,\legendmid+1.625) -- (12.85,\legendmid+2.125);
    \node [align = left, font = \scriptsize] at (13.4,\legendmid+1.875) {both};
    
        \node [align = center, font = \small] at (11,\legendmid+0.325) {difficulty of detection:};
    \filldraw [fill = target] (8.6,\legendmid-0.325) circle (5pt);
    \node [align = left, font = \scriptsize] at (9.4,\legendmid-0.325) {easy};
    \filldraw [fill = method] (10.6,\legendmid-0.325) circle (5pt);
    \node [align = left, font = \scriptsize] at (11.4,\legendmid-0.325) {harder};
    \filldraw [fill = concept] (12.6,\legendmid-0.325) circle (5pt);
    \node [align = left, font = \scriptsize] at (13.4,\legendmid-0.325) {hardest};

        \node [align = center, font = \small] at (11,\legendmid-1.575) {evaluation:};
    \node [align = center] at (9.35,\legendmid-2.225) {\Small\faCheck};
    \node [align = left, font = \scriptsize] at (10.15,\legendmid-2.225) {provided};
    \node [align = center] at (11.45,\legendmid-2.225) {\small\faQuestion};
    \node [align = left, font = \scriptsize] at (12.15,\legendmid-2.225) {missing};
\end{tikzpicture}%
}
        \caption{The bi-dimensional generalizability space. Axes represent dimensions along which to test generalization capabilities of detectors. The majority of existing detectors are evaluated under favorable conditions -- that is, only against a specific type of bots ($b_0$) and with data collected at a specific point in time ($t_0$) -- thus possibly overestimating their capabilities. The actual detection performance for $b \ne b_0$ and for $t > t_0$ are unknown. More realistic estimations could be obtained by evaluating detectors under more general conditions. Generalization along the $y$ axis can be achieved by adopting evaluation methodologies such as that proposed by Echeverr\'{i}a \textit{et al.}~\cite{de2018lobo}. Generalization along the $x$ axis can be obtained by applying adversarial approaches aimed at creating variations of currently existing bots.\label{fig:generalizability}}
\end{figure*}
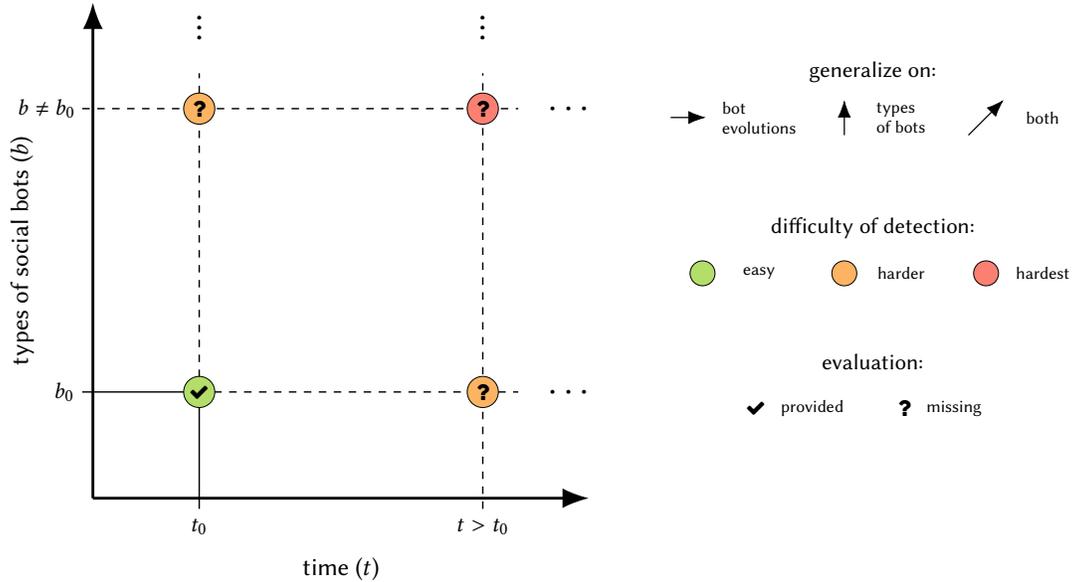
The exponentially growing body of work on social bot detection shown in Figure~\ref{fig:publications-trend}, somehow reassures us that much effort is bound to be devoted in the coming years to the fight of this crucial issue. However, at the same time it also poses some new challenges. Firstly, it is becoming increasingly important to organize this huge body of work. Doing so would not only contribute to a better exploitation of this knowledge, but would also allow researchers to more efficiently provide new solutions by avoiding exploring paths that already proved unsuccessful. To this end, this survey aims to provide a contribution to the critical review and analysis of the vast literature on and beyond this topic.

Secondly, the foreseen increase in publications inevitably implies that more bot detectors will be proposed. With the growing number of disparate detection techniques, it is becoming increasingly important to have standard means, such as benchmarks, frameworks and reference datasets, with which to evaluate and compare them. The present situation is one where we have a suitcase filled with all kinds of tools. Bad enough, we do not really know how to profitably use them, what are the differences between them, and ultimately, what they are really worth! Buying us yet another tool would not help much. Instead, a few targeted investments aimed at extensively evaluating and comparing our current tools, would tremendously increase the usefulness of all our suitcase.

One aspect that is often overlooked when evaluating bot detectors is their \textit{generalizability} -- that is, their capacity of achieving good detection results also for types of bots that have not been originally considered. To this regard, our analysis lays the foundations of a bi-dimensional generalizability space, sketched in Figure~\ref{fig:generalizability}. A desirable scenario for the near future would involve the possibility to evaluate any new bot detector against many different types of social bots, thus moving along the \textit{y} axis of Figure~\ref{fig:generalizability}, following the promising approaches recently developed in~\cite{de2018lobo,yang2019scalable}. It would also be profitable to evaluate detectors against different versions of current bots, thus somehow simulating the evolving characteristics of bots. This could be achieved by applying the adversarial approach previously described for creating many adversarial examples, opening up experimentation along the \textit{x} axis of the generalizability space. Combining these two evaluation dimensions, thus extensively exploring the generalizability space, would allow a much more reliable assessment of the detection capabilities of present and future techniques, thus avoiding overestimates of detection performance. In order to reach this ambitious goal, we must first create reference datasets that comprise several different kinds of malicious accounts, including social bots, cyborgs and political trolls, thus significantly adding to the sparse resources existing as of today\footnote{One of the few publicly available bot repositories is hosted at: \url{https://botometer.iuni.iu.edu/bot-repository/datasets.html}}. Here, challenges include the limited availability of data itself, missing or ambiguous ground truth and the obsolescence of existing datasets that hardly cope with the rapid evolution of malicious accounts. To this regard, continuous data-sharing initiatives such as that of Twitter for accounts involved in information operations\footnote{\url{https://transparency.twitter.com/en/information-operations.html}}, are extremely welcome as they can enable the next wave of research on these issues. Then, we should also devise additional ways for creating a broad array of diverse adversarial examples. Doing so would also require quantitative means to estimate the contributions brought by the different adversarial examples, for instance in terms of their novelty and diversity with respect to existing malicious accounts. These challenges currently stand as largely unsolved, and call for the highest effort of our scientific community.

Our longitudinal analysis of the first decade of research in social bot detection revealed some interesting trends. Early days were characterized by simple supervised detectors analyzing accounts \textit{individually}. Unsupervised detectors emerged in 2012-13 and shifted the target to \textit{groups} of misbehaving accounts. Finally, we highlighted the new rising trend of \textit{adversarial} approaches. Our analysis revealed that for more than a decade we fought each of the menaces posed by sophisticated social bots, cyborgs, trolls, and collusive humans, separately. Now, thanks to the rise of AI-enabled deception techniques such as \textit{deepfakes}, the most sophisticated of these malicious actors are bound to become indistinguishable from one another, and likely also from legitimate accounts. It is thus becoming increasingly necessary to focus on spotting the techniques used to deceive and to manipulate, rather than trying to classify individual accounts by their nature. Inauthentic coordination is an important piece of the deception puzzle, since it is exploited by bad actors for obtaining visibility and impact. Moreover, it is oblivious to the different types of bad actors.
In other words, both our findings and recent reflections~\cite{grimme2018changing,starbird2019disinformationspread} suggest that we should keep on moving away from simple supervised approaches focusing on individual accounts and producing binary labels. We should instead take on the challenging task of embracing the complexity of deception, manipulation and automation by devising unsupervised techniques for spotting suspicious coordination. In addition, future techniques should not provide oversimplistic binary labels, as often done  -- and as equally often criticized\footnote{\url{https://cyber.harvard.edu/news/2020-03/false-positive-problem-automatic-bot-detection-social-science-research}}, but should instead produce multifaceted measures of the extent of suspicious coordination. 
Our in-depth analysis revealed the emergence of group-based approaches several years before ``coordinated inauthentic behavior'' was acknowledged as the main threat to our online social ecosystems by the general public, and by the social platforms themselves. Among the most pressing challenges along this line of research is the problem of scalability of group-based detectors and the intrinsic fuzziness of ``inauthentic coordinaton''. In fact, the scalable and generalizable detection of coordination is still a largely open challenge, with only few contributions proposed so far~\cite{fazilsocialbots,pacheco2020uncovering}. Similarly, computational means to discriminate between authentic and inauthentic coordination are yet to be proposed and evaluated. Interestingly, the same analysis that anticipated worldwide interest in inauthentic coordination, is now suggesting that adversarial approaches might give us an edge in the long-lasting fight against online deception.

Summarizing the main suggestions stemming for our extensive analysis, future deception detection techniques should: (i) focus on identifying suspicious coordination independently of the nature of individual accounts, (ii) avoid providing binary labels in favor of fuzzier and multifaceted indicators, (iii) favor unsupervised/semi-supervised approaches over supervised ones, and (iv) account for adversaries by design. In addition, part of the massive efforts we dedicated to the task of detection should also be reallocated to measure (human) exposure to these phenomena and to quantify the impact they possibly have. Only through enacting these changes we will be able to develop tools that better represent the existing reality, thus providing actionable results to the many scientific communities and stakeholders looking at AI and Big Data tools as a compass to adventure in the perilous landscape of online information. These guiding lights stand in front of us as an exciting and rare opportunity, one that we did not have in the past. Acting upon and capitalizing on this opportunity is now exclusively on our shoulders.

\begin{acks}
Stefano Cresci is extremely thankful to all colleagues that collaborated with him on these topics throughout the years, for providing the invaluable ground on which this work is based. This research is supported in part by the EU H2020 Program under the scheme \texttt{INFRAIA-01-2018-2019: Research and Innovation action} grant agreement \#871042 \textit{SoBigData++: European Integrated Infrastructure for Social Mining and Big Data Analytics}.
\end{acks}

\bibliographystyle{ACM-Reference-Format}
\bibliography{bibliography}

\end{document}